\documentclass{article} 
\usepackage{iclr2024_conference,times}

\usepackage{amsmath,amsfonts,bm}

\def\eqref#1{equation~\ref{#1}}

\def\1{\bm{1}}

\DeclareMathAlphabet{\mathsfit}{\encodingdefault}{\sfdefault}{m}{sl}
\SetMathAlphabet{\mathsfit}{bold}{\encodingdefault}{\sfdefault}{bx}{n}

\usepackage{hyperref}
\usepackage{url}
\usepackage{soul, color, xcolor}
\soulregister{\cite}7 
\soulregister{\citep}7
\soulregister{\citet}7 
\soulregister{\ref}7 
\soulregister{\pageref}7 
\sethlcolor{yellow} 
\usepackage{graphicx} 
\usepackage{float} 
\usepackage{subfigure} 
\usepackage{colortbl} 

\title{Distributed Inference Performance Optimization for LLMs on CPUs}

\author{Pujiang He, Shan Zhou, Changqing Li, Wenhuan Huang, 
	 \AND Weifei Yu, Duyi Wang,  
	  Chen Meng \& Sheng Gui  \\
	  \\
	Datacener and AI Group \\
	Intel Corporation \\
	Shanghai, China \\
	\texttt{\{pujiang.he, shan.zhou, changqing.li,  wenhuan.huang\}}, \\
	\texttt{\{weifei.yu, duyi.wang, chen.meng, sheng.gui \}@intel.com} 
}

\iclrfinalcopy 
\begin{document}

\maketitle

\begin{abstract}

Large language models (LLMs) hold tremendous potential for addressing numerous real-world challenges, 
yet they typically demand significant computational resources and memory. Deploying LLMs onto a resource-limited hardware device with restricted memory capacity presents considerable challenges. 
Distributed computing emerges as a prevalent strategy to mitigate single-node memory constraints and expedite LLM inference performance.
To reduce the hardware limitation burden, we proposed an efficient distributed inference optimization solution for LLMs on CPUs. 
We conduct experiments with the proposed solution on 5th Gen Intel$^\circledR$ Xeon$^\circledR$ Scalable Processors, and the result shows the time per output token for the LLM with 72B parameter is 140 ms/token, 
much faster than the average human reading speed about 200ms per token.

\end{abstract}

\section{Introduction}
With the unprecedented success of Large Language Models (LLMs) across diverse domains (\citet{cui2023receive}, \citet{thirunavukarasu2023large}), 
the performance of LLM inference is paramount for extensive LLM applications (\citet{zhao2023survey}).
In the deployment of LLMs, we encounter numerous challenges, including substantial memory consumption, stringent latency targets, and long sequence lengths. Moreover, these challenges hindered the practical applications in low-resource environments.

As we known, LLMs primarily utilize the Transformers architecture (\citet{vaswani2017attention}), which exhibits high parallelism. However, efficiently deploying these models in practical applications presents challenges. 
This is because inference generation occurs token by token, with each token's computation relying on previously generated tokens. 
Multiple deployment optimization solution for LLM has been proposed, such as \citet{miao2023towards}, \citet{agrawal2023sarathi} and \citet{zheng2023response}. 
While these solutions are primarily designed for GPUs, when GPU hardware resources are limited, we can explore alternative options on CPUs.
Therefore, an efficient distributed solution for LLM on CPUs is proposed. 
A better distributed solution for LLM inference performance optimization on CPU is crucial for cost-savings, efficient hardware usage, and optimal inference strategies. It could help achieve the required scalability and efficient low-latency inference.

In this paper, we propose three approaches which helps optimize the distributed inference performance for LLMs on CPUs. We conduct experiments with the proposed solution on 5th Gen Intel$^\circledR$ Xeon$^\circledR$ Scalable Processors, and the results indicate that the LLM with 72B parameters achieves a time per output token of 140 ms, significantly surpassing the average human reading speed of approximately 200 ms per token.

\section{Approach}
\label{approach}

To optimize the distributed inference performance, minimizing communication cost wherever possible is important (\citet{bajovic2016distributed}). 
We utilize the oneAPI Collective CommunicationsLibrary (oneCCL) designed with the aim of creating a unified standard API compatible with various types of hardware accelerators.
Furthermore, we proposed multiple optimization approaches to enhance the LLM inference performance on CPUs as follows.

\subsection{Improve Scalability by Minimizing Synchronization}
\begin{figure}[h]
	\begin{center}
		\includegraphics[width=0.9\textwidth]{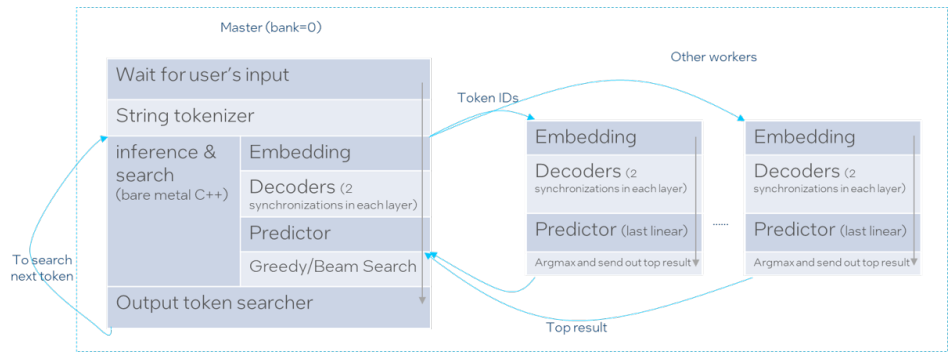}
	\end{center}
	\caption{Distributed inference based on oneCCL.}
	\label{dis1}
\end{figure}

During the initial phase of each inference round, the proposed solution broadcasts token IDs rather than broadcasting the values of Embedding part obtained based on token IDs. 
Similarly, we adopt an effective approach which is for each worker to compute top-k before performing the reduction at the end of each inference round. The implementation is shown in Figure ~\ref{dis1} .

\subsection{One-time Synchronization}
\begin{figure}[h]
	\begin{center}
		\includegraphics[width=0.9\textwidth]{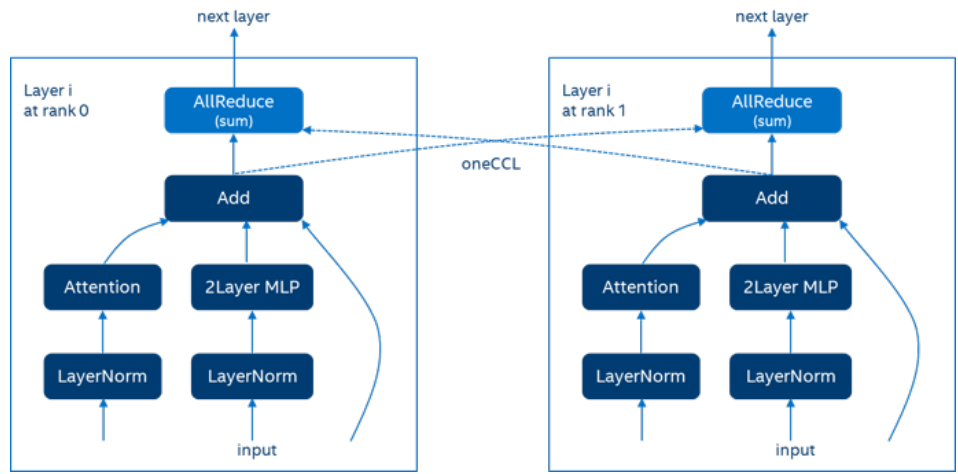}
	\end{center}
	\caption{One time synchronization.}
	\label{dis2}
\end{figure}

Optimizing communication cost based on each model's structure is essential. For models like GPT-J and Falcon, where attention and feed-forward network sections run in parallel, it's possible to achieve communication efficiency by ensuring that each decoder layer performs only one time synchronization which shows in Figure ~\ref{dis2} .

\subsection{Minimize Memory Copy}
As we are aware, when the computation module and communication module interact, data copying is often involved in practice. Therefore, a more aggressive optimization approach can be pursued to eliminate these copies.
This involves the computation module, during its last operation before communication, directly writing the results to the location of the communication module, achieving a zero-copy implementation.

\begin{figure}[h]
	\begin{center}
		\includegraphics[width=0.45\textwidth]{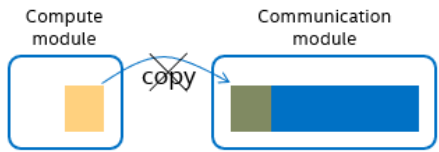}
	\end{center}
	\caption{Minimize memory copy.}
	\label{dis3}
\end{figure}

\section{Experiment Results}
\label{exp}

We conducted experiments using Qwen which is a large-scale pre-trained model developed by Alibaba Group (\citet{bai2023qwen}) with model parameter sizes of 72B. 
Qwen-72B is a Transformer-based large language model.
To illustrate performance results, we measured the per token time of the next token generation on the 4 * Intel$^\circledR$ Xeon$^\circledR$ Scalable Processors 8575C, where each device has 1 socket, and each socket has 48 cores.
With input size = 512 tokens and batch size = 1, the results shows time per output token is 140 ms/token, much faster than human reading speed.

\section{Conclusion}
We propose an efficient distributed inference performance optimization solution for LLMs on CPUs by leveraging oneCCL. 
The experiment results shows the promising per-token generation latency is 140 ms. 
In our future endeavors, we aim to enhance distributed LLM inference as a contribution to the open-source community. Furthermore, we intend to expand our approach to encompass a wider variety of CPUs, thereby empowering generative AI on CPUs in resource-limited environments.

\bibliography{iclr2024_conference}
\bibliographystyle{iclr2024_conference}

\end{document}